\documentclass[sigconf,nonacm]{acmart}

\usepackage{graphicx}
\usepackage{booktabs}
\usepackage{array}
\usepackage{ifthen}
\usepackage{subcaption}

\usepackage{algorithm} 
\usepackage{algpseudocode} 
\usepackage{listings}
\usepackage{xcolor}
\usepackage{amsmath}

\usepackage{amssymb}

\lstdefinestyle{cppstyle}{
  language=C++,
  basicstyle=\ttfamily\footnotesize,
  keywordstyle=\color{blue},
  commentstyle=\color{gray},
  stringstyle=\color{teal},
  numbers=left,
  numberstyle=\tiny\color{gray},
  stepnumber=1,
  numbersep=5pt,
  breaklines=true,
  showstringspaces=false,
  tabsize=2,
  captionpos=b
}
\AtBeginDocument{%
  }

\setcopyright{none}
\begin{document}

\title{QuietPrint: Protecting 3D Printers Against Acoustic
Side-Channel Attacks}


\author{Seyed Ali Ghazi Asgar}
\affiliation{%
  \institution{Texas A\&M University}
  \city{College Station, TX}
  \country{USA}}
\email{alighazi@tamu.edu}

\author{Narasimha Reddy}
\affiliation{%
  \institution{Texas A\&M University}
  \city{College Station, TX}
  \country{USA}}
\email{reddy@tamu.edu}



\renewcommand{\shortauthors}{Trovato et al.}

\begin{abstract}
The 3D printing market has experienced significant growth in recent years, with an estimated revenue of 15 billion USD for 2025. Cyber-attacks targeting the 3D printing process whether through the machine itself, the supply chain, or the fabricated components are becoming increasingly common. One major concern is intellectual property (IP) theft, where a malicious attacker gains access to the design file. One method for carrying out such theft is through side-channel attacks. In this work, we investigate the possibility of IP theft via acoustic side channels and propose a novel method to protect 3D printers against such attacks. The primary advantage of our approach is that it requires no additional hardware, such as large speakers or noise-canceling devices. Instead, it secures printed parts by minimal modifications to the G-code. 
\end{abstract}




\maketitle

\section{Introduction}
3D printing or additive manufacturing(AM), is a method of manufacturing components by depositing a specific material in a layer-by-layer manner. Over the years, due to the low cost of 3D printers, it has become accessible to a wider community, with an estimated revenue of more than 15 billion USD as of 2025 \cite{moneysonigara2025point,marketsandmarketsGlobalPrinting}. In addition, this printing technique is a suitable option for rapid prototyping objects with complex geometry while preserving quality \cite{ gibson2021additive,debroy2018additive,tofail2018additive}. As with every new technology that brings many benefits to the world, various risks also emerge. AM is also subject to many cyber security threats. One category of attacks on 3D printers focuses on damaging the printed component\cite{rais2025sabotaging,thakare2025secure}. In this scenario, the attacker tries to somehow manipulate the computer-aided design (CAD) file or the G-code file in a way that the manufactured part lacks necessary mechanical strength \cite{zeltmann2016manufacturing,rosselsecurity,rossel2023security}. For instance, authors in \cite{Trojan} showcased a type of attack in which an open-source firmware of a 3D printer is infected and  automatically reduces the strength of the component. Another type of attack directly focuses on the machine itself. For example, researchers in \cite{ahsan2025wattshield} simulated a scenario in which an attacker manipulates the firmware to bypass safety features that protect the nozzle from overheating beyond safe limits, causing fire hazards. Similar work by \cite{printGrave} demonstrates an attack called \textit{ Print Your Own Grave}, in which a malicious framework forces the 3D printer's nozzle onto the printer bed (which, in some models, is made of glass), causing the glass to break after 30 minutes of continuous pressure. Another issue in 3D printing domain is to manage multiple printers at the same site. Internet of Things (IoT) protocols, such as MQTT, can be used to monitor a 3D printing farms \cite{alshaheen2024rdm,garcia2023connecting}; however, this protocol has security weaknesses that could be exploited to damage the machines\cite{asgar2025analysis,mohammed2025enhancing}. In addition, designers are concerned about sharing their design files with manufacturers. Since malicious insiders and worker-related threats are potential vulnerabilities, several solutions have been proposed. One such method is to stream the G-code file directly to the machine by the designer \cite{baumann2017modelstream,tiwari2020cybersecuritystream}. Although this approach eliminates insider threats, it introduces significant risks to the manufacturing machine. If an attacker intercepts the streaming process and begins injecting malicious G-code commands during transmission, it can result in damage to the AM machine. This issue is addressed in the work proposed by \cite{asgar2025trustmanufacturertrustclient}, where instead of streaming the G-code file, the STL file is first split into chunks and then streamed sequentially. Since these STL fragments contain only inert design data and no machine commands, this method also mitigates the manufacturer's concerns.

\subsection{Previous Works and Side-Channel Leakage}

Prior research has shown that 3D printers are susceptible to a wide range of side-channel attacks that can compromise the confidentiality of design files. In these attacks, adversaries exploit indirect physical signals emitted during the printing process to infer motion patterns and ultimately reconstruct the printed object or its underlying G-code.

One prominent class of side-channel leakage arises from optical signals. Many modern 3D printers are equipped with cameras to enable remote monitoring or automated failure detection. While these cameras improve usability and reliability, they also introduce privacy risks if compromised. Previous work has demonstrated that visual data captured during the printing process can be analyzed to reconstruct the executed G-code, leading to intellectual property leakage \cite{liang2022hiding,chattopadhyay2025oneoptical}.

Power-based side channels have also been shown to leak sensitive information. By measuring the current drawn by stepper motors, researchers have been able to infer motion states such as starts, stops, and directional changes. These inferred states can then be synchronized across multiple axes to reconstruct movement trajectories and recover the original G-code \cite{power_attack}. Although power side-channel information has primarily been studied from an attack perspective, it can also be repurposed for defensive applications. Deviations between expected and observed current patterns may indicate anomalous or malicious firmware behavior.

Mechanical vibrations generated by stepper motors represent another source of side-channel leakage. Prior studies attached accelerometers to the printer chassis and analyzed vibration spectrograms to identify strong correlations between vibration patterns and nozzle movements. Using this information, researchers successfully reconstructed printed geometries, highlighting the sensitivity of vibration signals to printer motion \cite{vibration}.

Finally, magnetic side-channel attacks exploit the electromagnetic fields produced by energized stepper motor coils. As the motor phases are activated sequentially, they induce measurable variations in the surrounding magnetic field. Several studies have demonstrated that these magnetic emissions can be captured using commodity sensors, such as those embedded in smartphones placed near the printer, and subsequently used to reconstruct printed objects \cite{magneticjamarani2025practitioner,magneticsong2016my}.

\subsection{Acoustic side-channel}
When a 3D printer operates, it generates acoustic noise from several sources, with stepper motors and nozzle fans being the primary contributors \cite{bayens2017see}. In the work conducted by \cite{acoustic-chhetri2017confidentiality,acoustical2016acoustic}, a professional audio recorder (Zoom H6) was placed near a 3D printer. By extracting audio features and training multiple machine learning models to predict the speed and direction of movement, the researchers demonstrated that it is possible to reconstruct the G-code using audio signals. In another study \cite{acoustickubiak2020usefulness}, two professional microphones (Behringer C-1U) were placed on the left and right sides of the 3D printer. Then, the spectrograms of the audio from both channels were further processed for reconstruction. To address this problem, \cite{noise_acoustic_stanczak2023evaluation} proposed adding a speaker with an appropriate amplifier placed behind the printer to generate enough noise such that the signal-to-noise ratio (SNR) of the recorded sound becomes unrecognizable. Although this method is effective, it is costly, with an estimated expense of approximately €1000 and requires significant space.

The main focus of this work is to protect 3D printers against acoustic side-channel attacks. Although multiple proof-of-concept attack scenarios have been proposed \cite{acoustic-chhetri2017confidentiality,acoustical2016acoustic,acoustickubiak2020usefulness}, to the best of our knowledge, no viable solution has been proposed yet to protect 3D printers from this category of attacks. In this work, we first analyze the different sources of acoustic noise in 3D printers. Then, we evaluate the effectiveness of each acoustic source and its contribution to the reconstruction of the printed object. Finally, we will provide defense mechanisms to obfuscate the acoustic information in a way that it is no longer reconstructible. 


The remainder of the paper is organized as follows. We first review the technical terminology in Section \ref{Terminology}. Next, we present our threat model in Section \ref{threat_model}, followed by our hardware details. We then provide a proof of concept approach to reconstruct location from audio in Section \ref{sec:reconstruct}. Subsequently, we will explore different defense methods in Section \ref{sec:defense} and evaluate its performance. Finally, the  conclusion is discussed in Section  \ref{sec:conclusion}.

\section{Terminology used in additive manufacturing}
\label{Terminology}

Designers utilize 3D CAD applications to translate conceptual ideas into  digital models. Since each CAD software employs its own proprietary file format, a standardized format is essential to ensure compatibility across various manufacturing systems. As the name suggests, the Standard Triangle Language (STL) file converts the original CAD model into a mesh of triangles used to reconstruct the entire object. Each triangle contains specific coordinates and geometric properties. 3D printers cannot directly execute STL files. An STL file is first sliced with a slicer application and the trajectory of the printing process is specified sequentially in terms of G-code commands. Each G-code command serves a specific purpose once executed on the machine. For instance, "\texttt{G0 X6}" moves the nozzle to the location X= 6 mm along the X-axis without  extruding filament or the command "\texttt{M106 S100}" can set the nozzle fan speed to 100\% (See Table\ref{table:gcode_table}). 

\begin{table}[b]
\centering

\caption{Example of G-Code  commands}
\label{table:gcode_table}

  \begin{tabular}{| >{\centering\arraybackslash} m{1.2cm}| >{\centering\arraybackslash} m{2cm} | >{\centering\arraybackslash} m{4cm} |}
  
    \hline
 Command & Example & Description \\
    \hline
 G1 & G1 X1.0 Y3.3 & Move to the location 1.0mm on X axis and 3.3mm on the Y axis with extrusion on\\

\hline
 
 G28 & G28 X Y & Home X and Y axis \\

\hline
M106 & M106 S50 & Set the fan speed to 50 \% \\
\hline
M104 & M104 S235 & Set the the hotend temperature to 235 $^{\circ}$C \\
\hline
M997 & M997 & Update the firmware from SD card \\
\hline
M999 & M999 & Restart the machine \\
\hline

 \end{tabular}

\end{table}

\section{Threat Model}
\label{threat_model}
In this work, we focus on a scenario in which a device with audio recording capabilities is located near the 3D printer. This device may be compromised, allowing an attacker to gain access to its microphone, or it may have been intentionally placed by a malicious insider, such as a worker . In this work, we focus on defending 3D printers against acoustic leakage, ensuring that even if  attackers records the audio signal, they cannot reconstruct the original printed model (e.g., the G-code).

\begin{figure}[!t]
    \centering
    \includegraphics[width=1\linewidth]{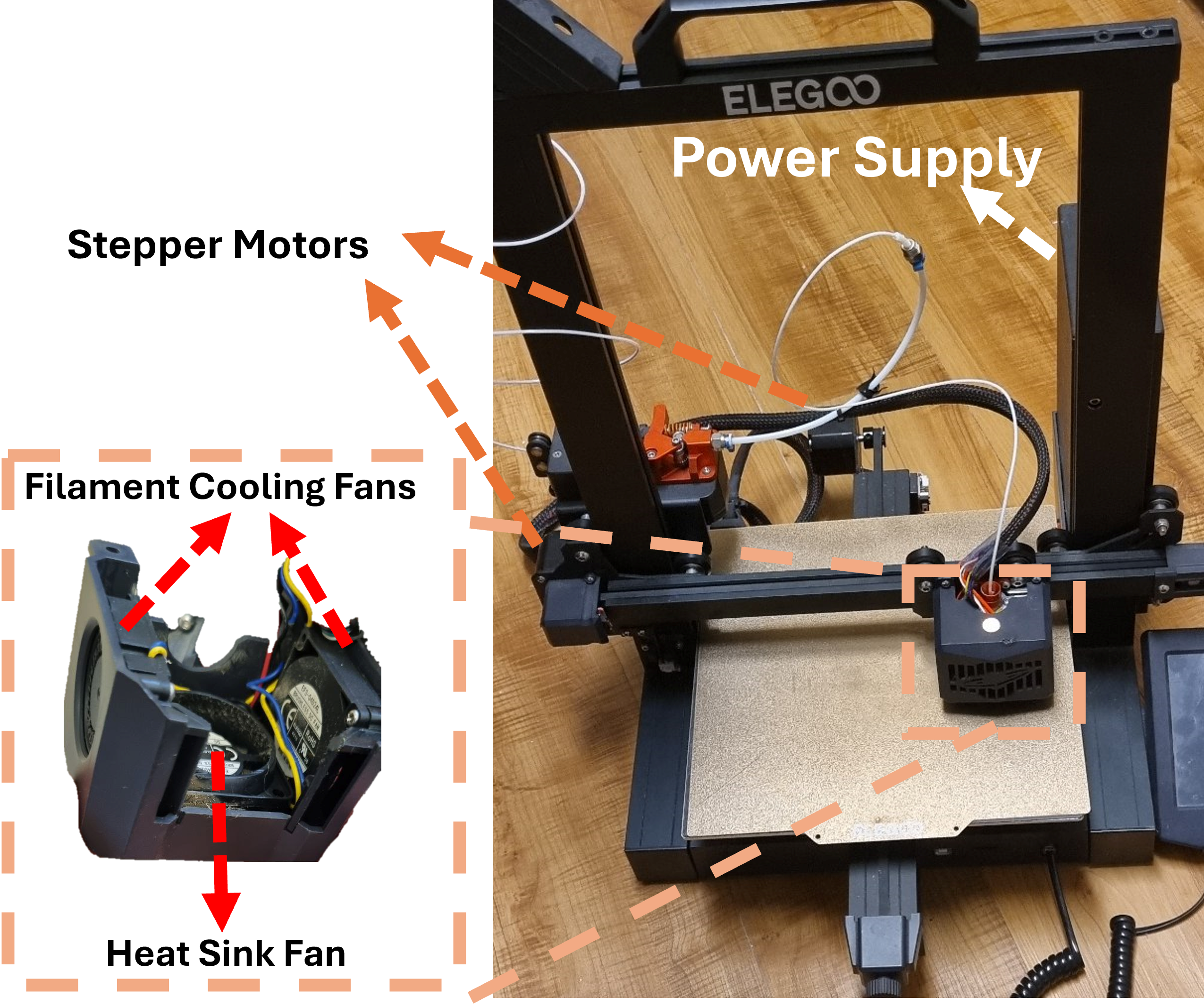}
    \caption{Sources of acoustic noise. }
    \label{fig:acoustic_source}
\end{figure}

\section{Data Acquisition Setup}
\label{sec:hardware}
\subsection{Hardware }
We used an Elegoo Neptune 3 FDM 3D printer as our test case for data collection and printing test objects. For audio recording, to simulate a real-world scenario, we utilized the built-in microphone of a Microsoft Surface Pro 7+ laptop  while some previous studies relied on professional microphones which may be impractical in realistic attack scenarios.

\subsection{Sources of Acoustic Information}
\subsubsection{Power Supply} As shown in Figure \ref{fig:acoustic_source}, there are three main sources of acoustic noise. The first, which contains almost no useful information, is the power supply cooling fan. This fan operates to dissipate heat generated by the power supply circuit and is not correlated with G-code commands and movements. 

\subsubsection{Stepper motors} Stepper motors operate based on the principles of electromagnetism. A magnetic core, known as the rotor, is placed at the center and surrounded by multiple coils. When a coil is energized, the rotor rotates toward it due to magnetic attraction. By energizing the coils in a specific sequence, the rotor performs stepping movements which also generate a significant amount of acoustic noise \cite{acoustical2016acoustic}.
Each axis of a 3D printer is controlled by a separate stepper motor. As shown in Figure \ref{fig:acoustic_source}, the X and Y axes are each driven by their own stepper motors. Therefore, by analyzing patterns in the recorded audio and isolating the acoustic features of each motor, it becomes possible to estimate the rotational speed and  convert that speed into the corresponding linear distance.

\subsubsection{Cooling Fans} Another source of audio is generated from the fans placed on top of the 3D printer's nozzle. As shown in Figure \ref{fig:acoustic_source}, there are usually three fans placed on the nozzle. Two of these fans are \textit{Filament cooling fans} that are placed on the sides, and their outlet is toward the 3D printer's bed. The main purpose of these fans is to cool the melted filament. The heat sink fan is also located on the extruder and helps preventing the meltdown of the filament inside the extrusion tube. 

\begin{figure}[!b]
    \centering
    \includegraphics[width=1\linewidth]{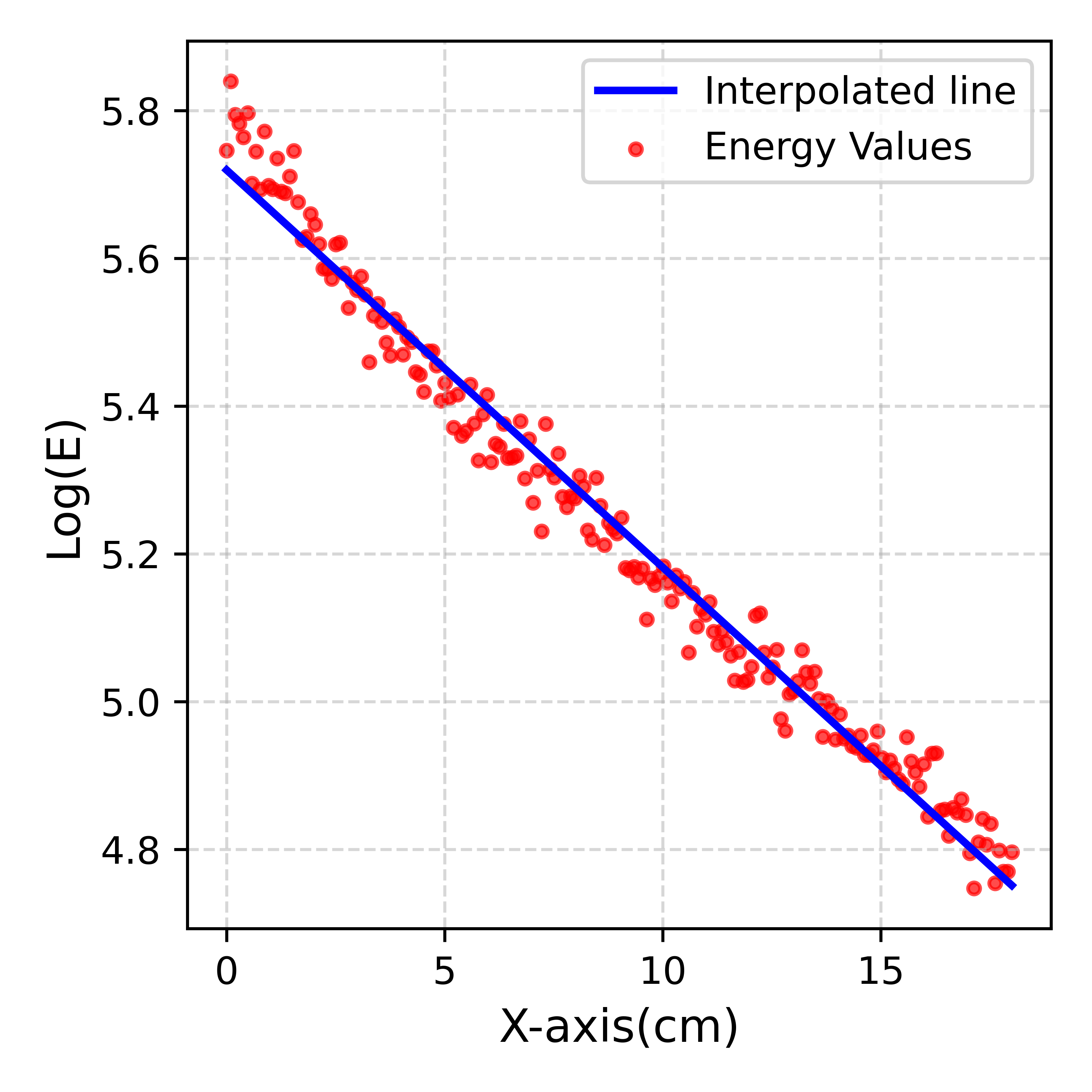}
    \caption{Linear interpolation of audio energy to horizontal motion. }
    \label{fig:linear fit}
\end{figure}

To investigate the effect of important acoustic features on information leakage. The first question that arises is whether it is possible to reconstruct the nozzle's location using only the noise generated by the fan. In our 3D printer setup, the nozzle moves along the X-axis, while the bed moves along the Y-axis. Therefore, we focus on analyzing whether the fan noise alone can be used to infer the nozzle’s movement along the X-axis. 
To achieve this goal, we designed an experiment. First, we moved the nozzle from x = 0 cm to x = 18 cm with the speed of 500 mm/min.We applied a fourth order high-pass Butterworth filter. Then we calculated the energy of the audio for every 100~ms splits according to the following formula (we used sum of absolute values for lower complexity instead of sum of squares):
\[E = \frac{1}{N} \sum_{i=1}^{N} |x_i|\]
and then we used the $\log(E)$ for interpolating a single line which shows that the $\log(E)$ has a linear correlation with the horizontal movement (See Figure \ref{fig:linear fit}).

\begin{figure*}[!t]
    \centering
    \includegraphics[width=1\linewidth]{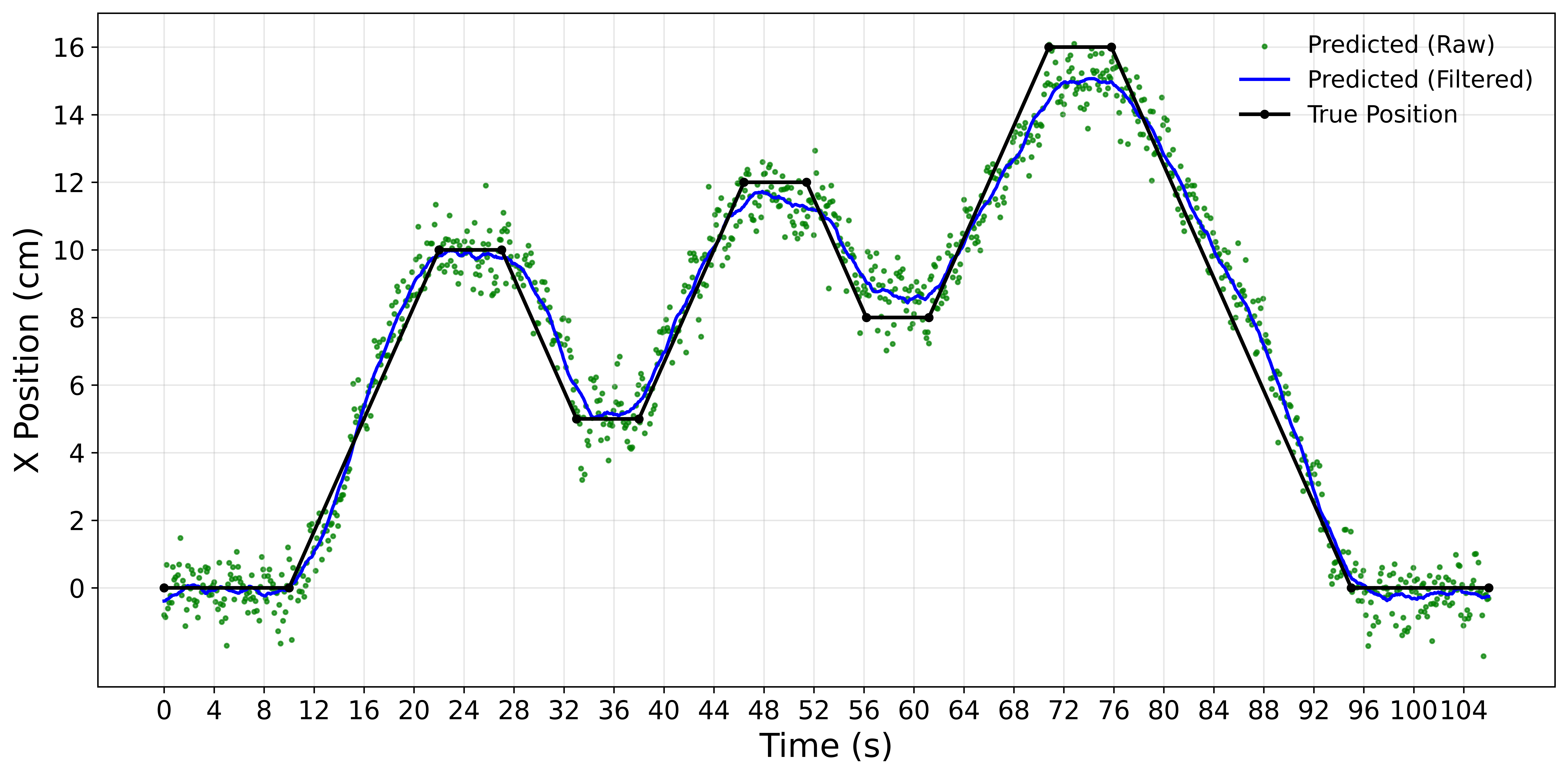}
    \caption{Predicting the nozzle motion along the X-axis using randomized movements.}
    \label{fig:location_predictoin_single_axis}
\end{figure*}

Next, we used the interpolated line for our next experiment which is a custom G-code sequence that first moves the nozzle to X = 0 cm and holds that position for 10 seconds. It then moves to X = 10 cm, followed by X = 5 cm. Next, it moves to X = 12 cm, waits briefly, proceeds to X = 8 cm, then to X = 16 cm, and finally returns to X = 0 cm. At each step, the nozzle pauses for 5 seconds. The movement speed is set to F500, which corresponds to 500 mm/min. As it is shown in Figure \ref{fig:location_predictoin_single_axis}, attacker can easily predict the location of the nozzle only using a single feature(energy) with a simple interpolated line that we got from the previous step (moving from X = 0~cm to X = 18~cm). Therefore, if the attacker can estimate the correct interpolation line based on the energy of the audio generated by the extruder fan, he may be able to accurately pinpoint the current location of the nozzle. Even more concerning, in the case of a CoreXY printer—where the nozzle can move along both the X and Y axes—an attacker with access to two audio recording devices could potentially reconstruct the nozzle's position deterministically in both axes.

\subsection{Synchronizing Location with Audio}
\label{sec:sync}

The main goal of an acoustic side-channel attack is to reconstruct nozzle locations from recorded audio. To accomplish this, a model is needed that takes audio as input and predicts the corresponding locations. This requires knowing which point in the audio aligns with the nozzle’s position at any given time.

In a recent study \cite{datasetmadamopoulos20243d}, researchers released a dataset containing audio recordings of 3D printed objects. However, a key limitation of this dataset is the lack of synchronization between the audio and the nozzle positions. As a result, it becomes challenging for the research community to train machine learning models that map audio to precise locations due to the absence of time-aligned data.

To address this issue, we developed a custom synchronization method, provided in Algorithm \ref{alg:sync}. We first set up a serial connection to the 3D printer. At the beginning of the process, we simultaneously start audio recording and initialize a timer. Then, we send the first G-code command to the printer. After the printer receives the first G-code line, we take advantage of the \texttt{M400} command. This command ensures that no subsequent movement commands will be executed until the current motion is completed. By extracting the coordinates from the previous G-code line, we can  determine the current nozzle position. We then record the nozzle's coordinates along with the current timestamp. Since the audio recording and timer were started simultaneously, we achieve consistent timing between the location data and the recorded audio. The final output of our data acquisition process consists of nozzle locations, time points, and recorded audio.

\begin{algorithm}[!t]
	\caption{Audio and Location Synchronization} \label{alg:sync}

	\begin{algorithmic}[1]
            \State Initialize serial connection
            \State Set audio recording parameters
            \State $\textit{locations} \gets [\,]$
            \State $\textit{times} \gets [\,]$

            \Function{send\_gcode\_to\_printer}{\textit{gcode\_file\_path}}
                    \State  Open G-code file
                    \State  Start audio record and timer
                    \ForAll{line in file}
                        
                        \If{$\textit{line} = \texttt{''}$ or $\textit{line}[0] = \texttt{';'}$}
                            \State \textbf{continue}
                        \EndIf
                        
                        \State  Send \textit{line} and wait for "ok"
                        \State  Send \texttt{"M400"} and wait for "ok"
  
                        \State  Extract x, y, z using \texttt{extract\_xyz\_Gcode}
                        \State  Get current \texttt{elapsed\_time}

                        \If{$x \neq \texttt{None}$ or $y \neq \texttt{None}$ or $z \neq \texttt{None}$}
                            \State  Append \texttt{[x,y,z]} to \texttt{locations}
                            \State  Append \texttt{elapsed\_time} to \texttt{times}

                        \EndIf
                        
                    \EndFor

            \EndFunction

            \State Call \texttt{send\_gcode\_to\_printer(filename)}
            \State Call \texttt{stop\_audio\_recoding()}
            \State Save $(locations, times)$ to 'loc\_time.pkl'

    \end{algorithmic} 
\end{algorithm}

\begin{figure}[!b]
    \centering
    \includegraphics[width=0.85\linewidth]{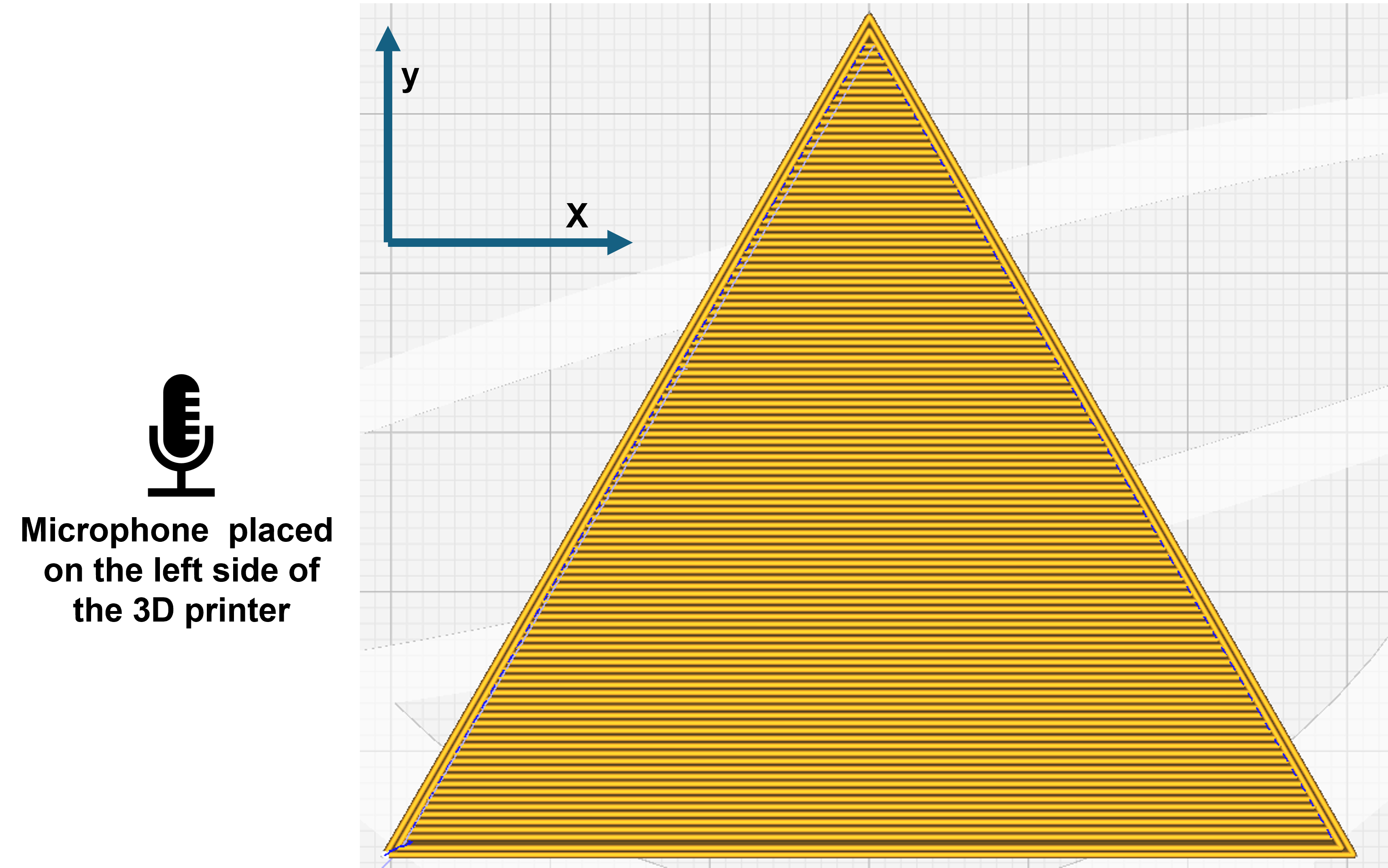}
    \caption{G-code of the test object with horizontal slicing angle and microphone's position. }
    \label{fig:test_trainagle}
\end{figure}

\begin{figure*}[!t]
    \centering
    \includegraphics[width=1\linewidth]{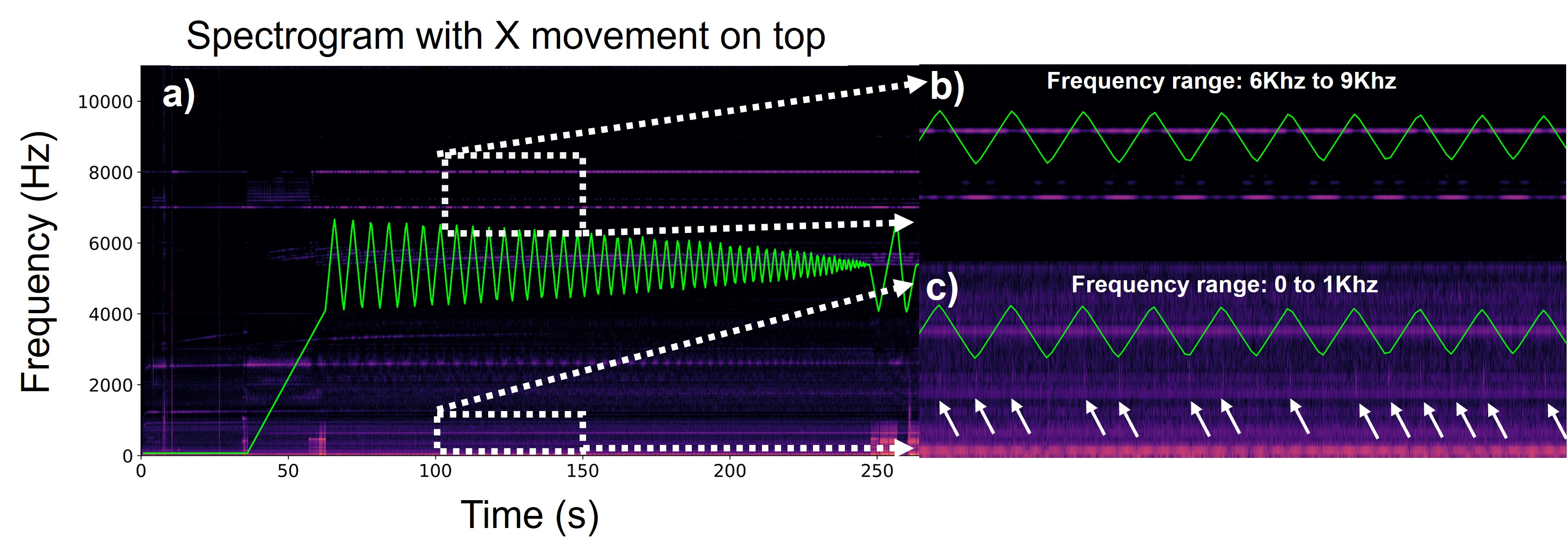}
    \caption{(a) Nozzle positions along the X-axis overlaid on the spectrogram of the corresponding audio. (b) Zoomed-in view of the time range between 100 and 150 seconds, and the frequency range from 6 kHz to 9 kHz. (c) Zoomed-in view of the time range between 100 and 150 seconds, and the frequency range from 0 to 1 kHz. Nozzle position along the X-axis is shown in green.  }
    \label{fig:sync_loc_audio}
\end{figure*}

To evaluate the final results, we designed a test object as shown in Figure \ref{fig:test_trainagle}. The slicing angle was set to 90 degrees, resulting in horizontal tool-paths along the X-axis. The laptop used for audio recording was placed on the left side of the 3D printer. Hence, we expect to observe higher audio volumes when the nozzle is at its leftmost position (closest to the microphone), and lower volumes when it is at the rightmost position (farthest from the microphone).We provide the spectrogram of the recorded audio, along with the nozzle's X-axis positions (plotted in green), in Figure \ref{fig:sync_loc_audio}. For better visibility, middle-range frequencies that did not show noticeable correlation with nozzle movement have been omitted. Based on our observations, two significant patterns emerge that support the effectiveness of our synchronization method. The first pattern appears in the frequency range between 6 kHz and 9 kHz, originating from the nozzle’s fan. As shown in Figure \ref{fig:sync_loc_audio}(b), when the nozzle is at its leftmost position (nearest to the microphone), the spectrogram shows higher energy levels. Conversely, when the nozzle is at its rightmost position, the energy levels are lowest. In addition, we observed a second pattern in the frequency range between 0 and 1 kHz. This audio pattern is generated by sudden changes in the nozzle’s path. For example, when the nozzle reaches the far right side and quickly reverses direction to start a new line, it produces a distinct movement sound. These sharp peaks, indicated by white arrows in the figure, are synchronized with the changes in the nozzle’s direction.

\section{Reconstructing Location from Audio}
\label{sec:reconstruct}

\begin{figure}[!b]
    \centering
    \includegraphics[width=1\linewidth]{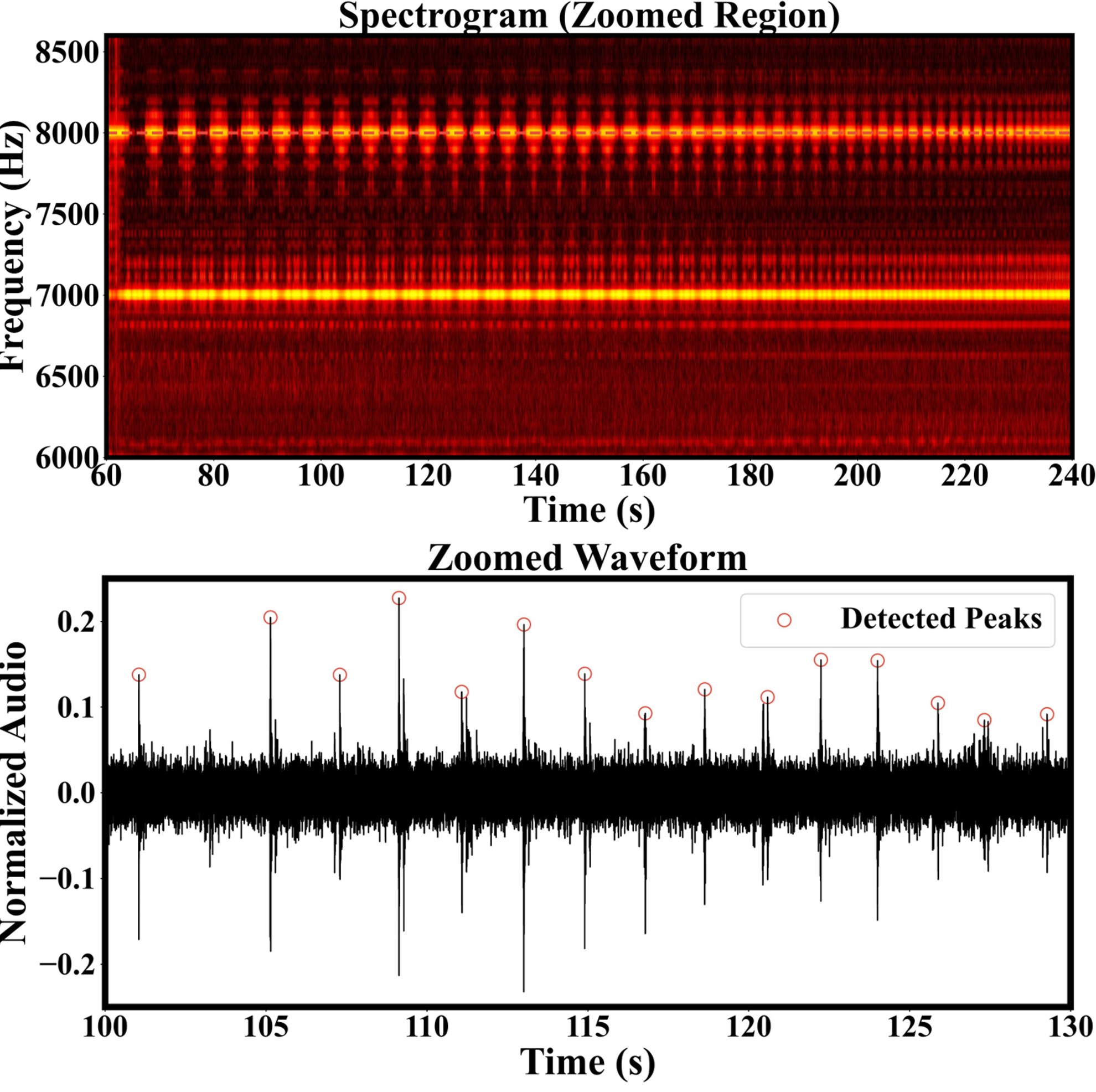}
    \caption{The spectrogram shows the noise generated by the fans, with a dominant frequency component at 8,000 Hz. The temporal plots display sudden spikes produced by the stepper motors. These peaks occur when the motors change direction and are identified using a peak detection algorithm.}
    \label{fig:audioPeak}
\end{figure}

\begin{figure}[!b]
    \centering
    \includegraphics[width=1\linewidth]{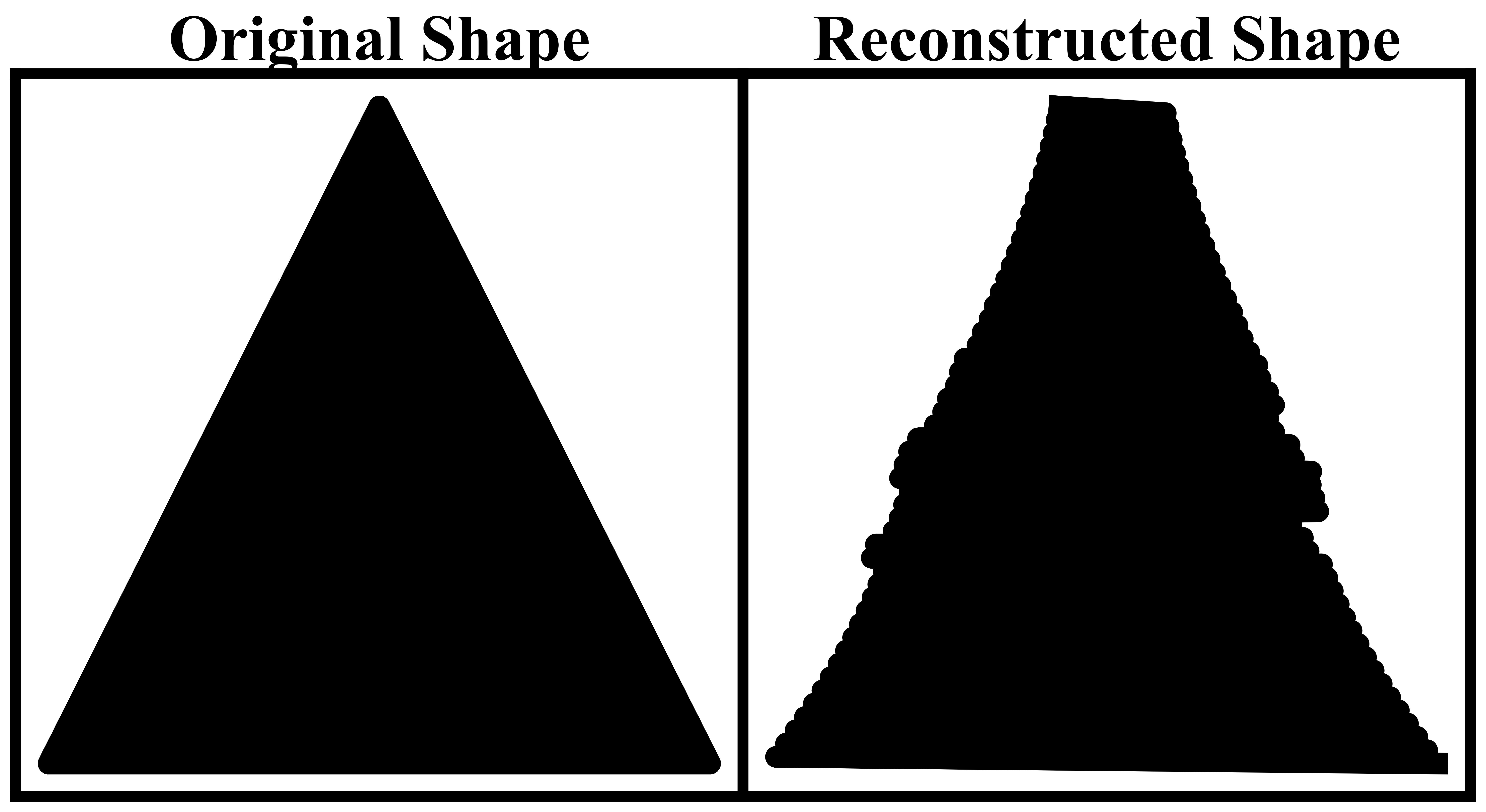}
    \caption{The real shape is shown on the left and the predicted shape based on the sudden movements from stepper motors is shown on the right.  }
    \label{fig:sudden_pred}
\end{figure}

One method to reconstruct locations from an audio signal is through the noise generated by the stepper motors. When the stepper motors reach their final destination and begin a new motion, they produce a sudden spike due to the abrupt change in direction (See Figure \ref{fig:audioPeak}). Assuming that the stepper motor operates at a constant printing speed, the time difference between two consecutive points can be used to estimate the distance traveled. By multiplying the time difference by the speed, we can calculate the distance covered. Combining all these distances allows us to reconstruct the overall shape.
To demonstrate that it is possible to reconstruct the shape based on these spikes, we reconstructed the data points along the X-axis. We assume that the Y-axis movement increments at a constant distance due to horizontal slicing. First, we filtered the frequency range of the audio signal between 100 and 600 Hz and then amplified the signal. This filtered audio signal is shown in Figure~\ref{fig:audioPeak}. Next, we normalized the data and identified the peaks within the acceptable range. Some of these peaks corresponded to environmental noise rather than sudden movements, so we ignored them. We then used the time differences between consecutive changes in the path to calculate the length of motion. Specifically, we obtained the time difference between each of these peaks, applied a Savitzky–Golay filter, and further smoothed the data. Finally, we multiplied each time-difference value by a constant speed (1200~mm/min) and toggled the movement direction from left to right and right to left by changing the sign of the speed. The resulting X-axis location prediction, based on the sudden acoustic spikes from the stepper motors, is shown in Figure~\ref{fig:sudden_pred}. Similar to the sudden spikes of stepper motors, it is possible to reconstruct the object using the noise generated from the fan. Based on our experiment, filament cooling fans also generate a unique high-pitch (8~kHz) noise. The spectrogram of this noise is shown in Figure \ref{fig:audioPeak}.


\section{Defense Method}
\label{sec:defense}
To obfuscate the noise generated by the stepper motor and fans, we propose the concept of Stealth Head Movement (SHM). In this method, we extend the normal motion of the 3D printer's head in such a way that even if the attacker can decode the acoustic signal and recover the shape, the recovered shape will be different than the original one. 
As shown in Figure \ref{fig:approach2}, we first obtain the original G-code file and  bound the object within a rectangular boundary. Then, we extend all the motions to this maximum length. We find a point that is collinear with the previous path and extend the line until the shape remains within a rectangle defined by the maximum and minimum boundaries. Once the new point is added, a new G-code line is also added to the G-code file so that the nozzle will return to the next point. This approach has a couple of benefits. First of all, as shown in Figure \ref{fig:audioPeak}, when the nozzle changes direction, it may generate a sudden acoustic spike. However, with the proposed approach, since the nozzle path changes at the boundary of the rectangle, even if such sudden spikes occur, the attacker cannot recover the original shape's locations. Without this extension movement, though, if the attacker detects these spikes, he could reconstruct the original shape. Second, even if an adversary exploits acoustic emissions from cooling fans to infer location, they will recover the obfuscated shape rather than the original. To further investigate this method, there are two main research questions. The first question is whether, at the transition points -- where the extension move is added to the G-code commands -- there are any recognizable audio features that an attacker can detect to identify the transition and distinguish between the original motion and the extended motion.
The second question is how to remove some of these extended movements to optimize the print time and eliminate unnecessary extra motions. We first demonstrate the effectiveness of our method, and then we will investigate these questions in the following subsections.

\subsection{Distinguishable audio pattern at transition points}
To evaluate the effect of our approach on the audio pattern, we first split our spectrogram at the start and end points of the bounded rectangle (See Figure \ref{fig:approach2}(a)) which are the leftmost and rightmost points. Then, we pinpoint the locations at which the transition between normal print and extended print is taking place. We visually inspected different segments, but there was not any recognizable pattern that can distinguish the transition points, from its surroundings.

\begin{figure}[!t]
    \centering
    \includegraphics[width=1\linewidth]{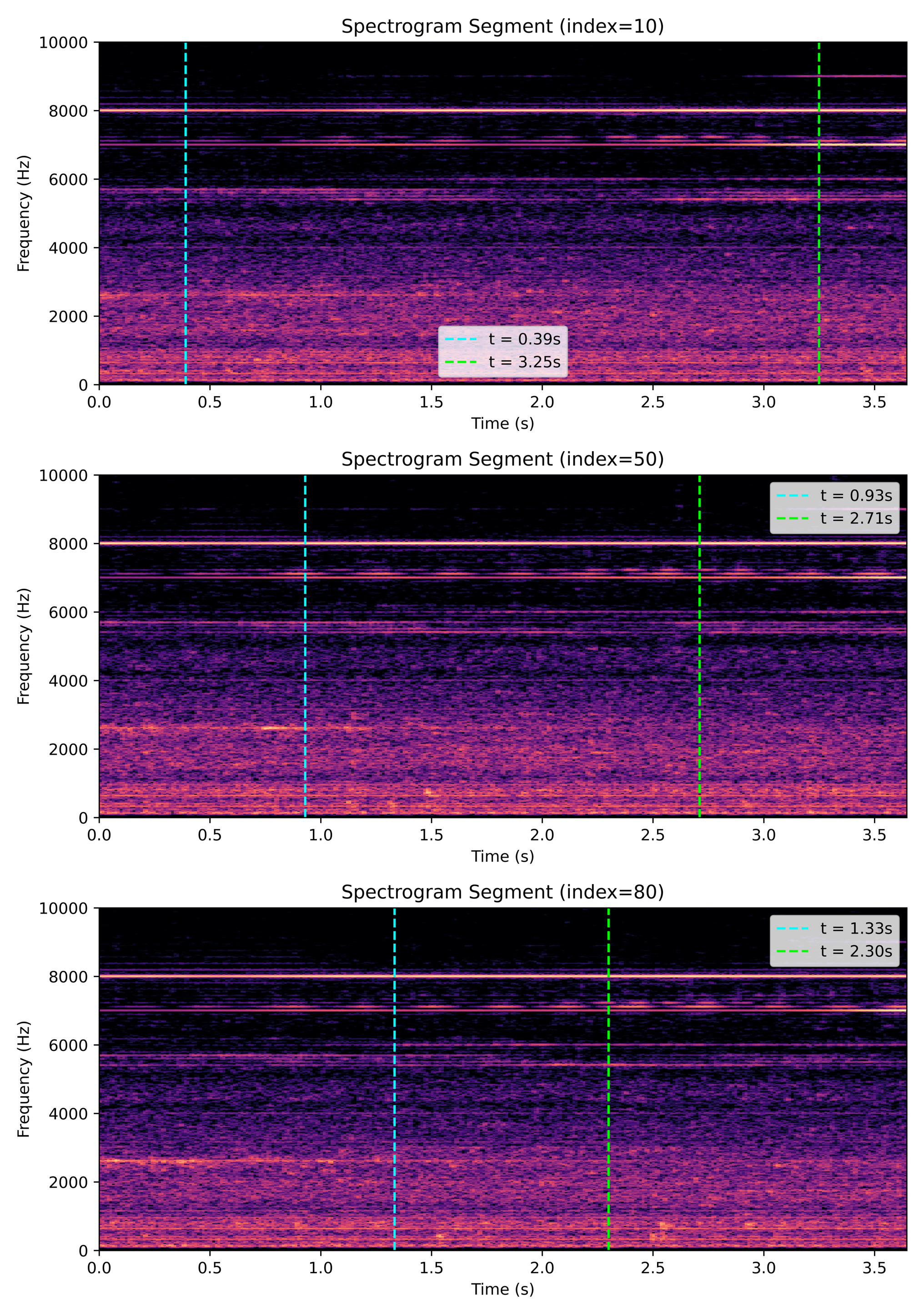}
    \caption{Spectrogram of different segments and the locations at which transition between normal path and extended path are taking place.}
    \label{fig:spec_trans}
\end{figure}

To further validate our claim, we resized all the segments to the same size, flattened the spectrograms, and trained an XGBoost regression model to predict the transition locations. We used 300 segments, using 30\% as the test set without shuffling to avoid data leakage. The model showed a high mean absolute percentage error of 33.79\% and a low Pearson correlation coefficient of 0.4. These results suggest that it is not feasible for an attacker to identify the transition points(See Figure\ref{fig:spec_trans}). 

\subsection{Optimizing extended paths}

    
\begin{figure}[!b]
    \centering
    \includegraphics[width=1\linewidth]{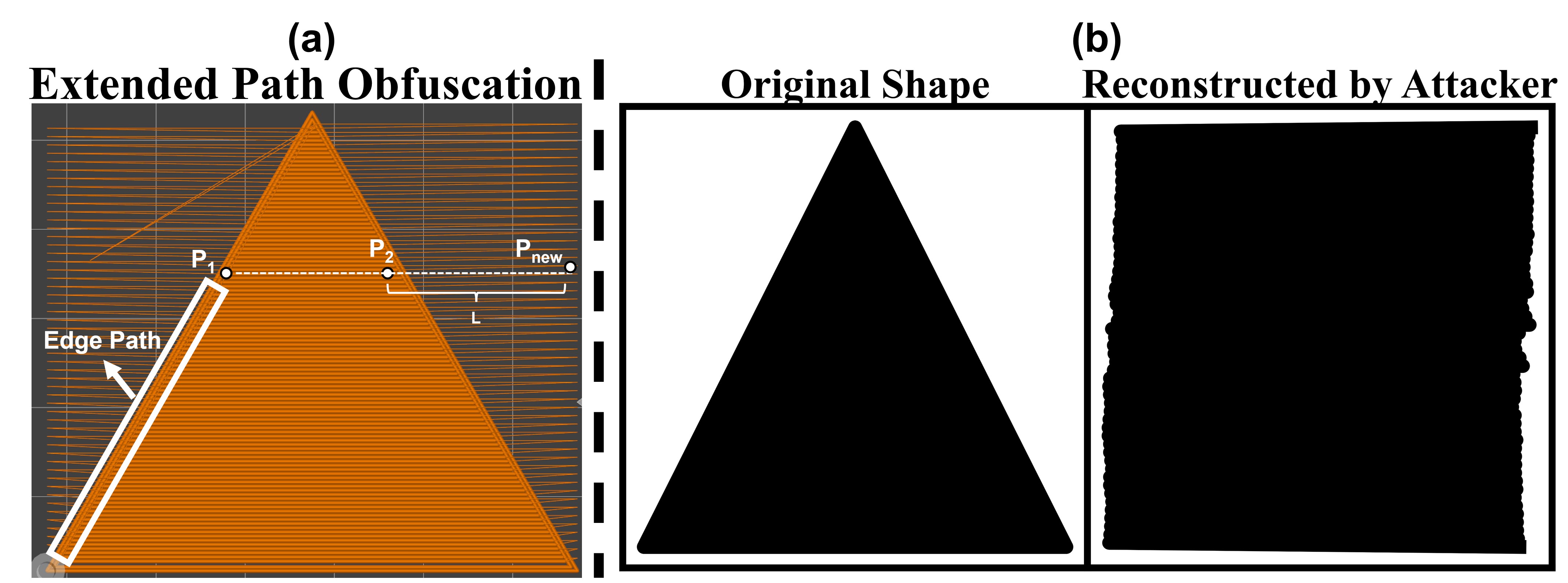}
    \caption{(a) Extending each path to the boundary of a rectangle by adding a collinear point along the same direction to obfuscate the object. (b) Original triangle is shown on the left and what attacker deciphered from audio file is shown on the right.}
    \label{fig:approach2}
\end{figure}

\begin{figure*}[!t]
    \centering
    
    \includegraphics[width=1\linewidth]{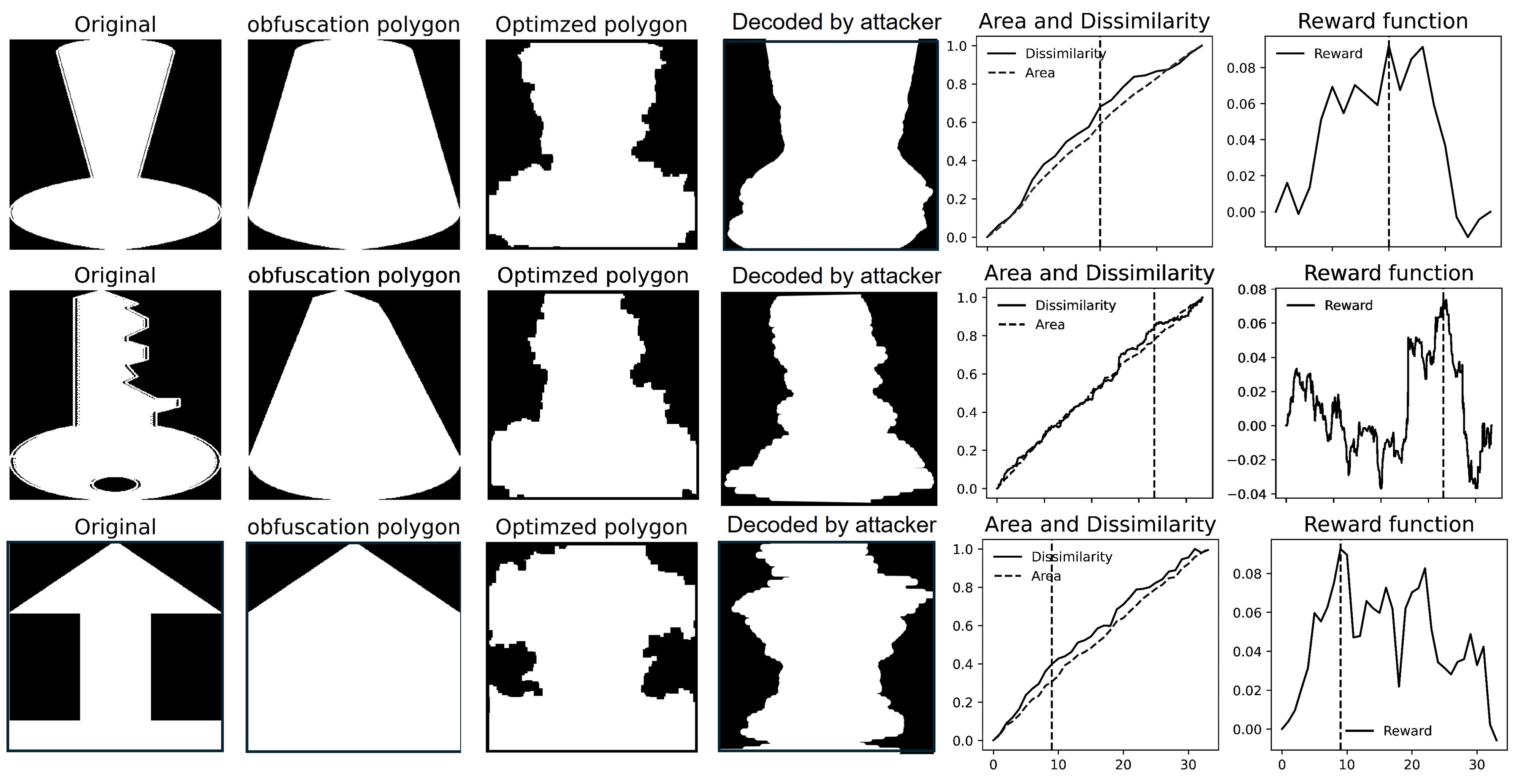}
    \caption{The figure shows the results of the obfuscation algorithm. The first column presents the original shape, followed by the naive obfuscation and the optimized obfuscation polygon. The fourth column shows the shape an attacker reconstructs from recorded audio. The optimized polygon’s area, dissimilarity, and reward value are also shown.}
    \label{fig:before_after}
\end{figure*}

Although bounding a shape within the rectangle obfuscates the shape, it might not be the optimized solution. For instance, in the case of the triangle shown in Figure \ref{fig:approach2}(a) since the area of the extended paths region is  same as the triangle, the printing time triples (the area of the extended regions is same as  the triangle, and for each movement to the edge there is another return motion). Hence, it is necessary to further optimize the extended paths region. 
As shown in Algorithm \ref{alg:apprhac2alg}, we first fit a convex hull around the object. The goal is to remove parts of the convex hull to minimize the area difference between the optimized shape and the original shape. However, the dissimilarity between the original and optimized shapes must remain high; otherwise, an attacker may be able to reconstruct the original shape. We can formulate our problem as following:
\[
 R(\mathbf{x}) = D(\mathbf{x}) - A(\mathbf{x})
\]
and the index of the optimized shape can be found as:
\[
\arg\max_{\mathbf{x}} \; R(\mathbf{x} )
\]
Where \( D \) represents the dissimilarity between the optimized shape and the original shape, and \( A \) denotes the number of added white points in the optimized shape.
We selected Procrustes analysis as the metric to measure dissimilarity between two shapes. This algorithm aligns shape \( Z \) with shape \( W \) using linear transformations such as scaling, rotation, and translation \cite{gower2010procrustes,krzanowski2000principles} and reports a dissimilarity metric which higher values indicate greater differences between the two shapes. Both matrices \( Z \) and \( W \) have the same size and by applying linear transformation to the second matrix, our goal is to minimize the following value:
\[
\text{D} = \sum_{i=1}^{n} ( \mathbf{Z}_i - \mathbf{T}(\mathbf{W}_i) )^2,
\]

\begin{figure}[!b]
    \centering
    \includegraphics[width=1\linewidth]{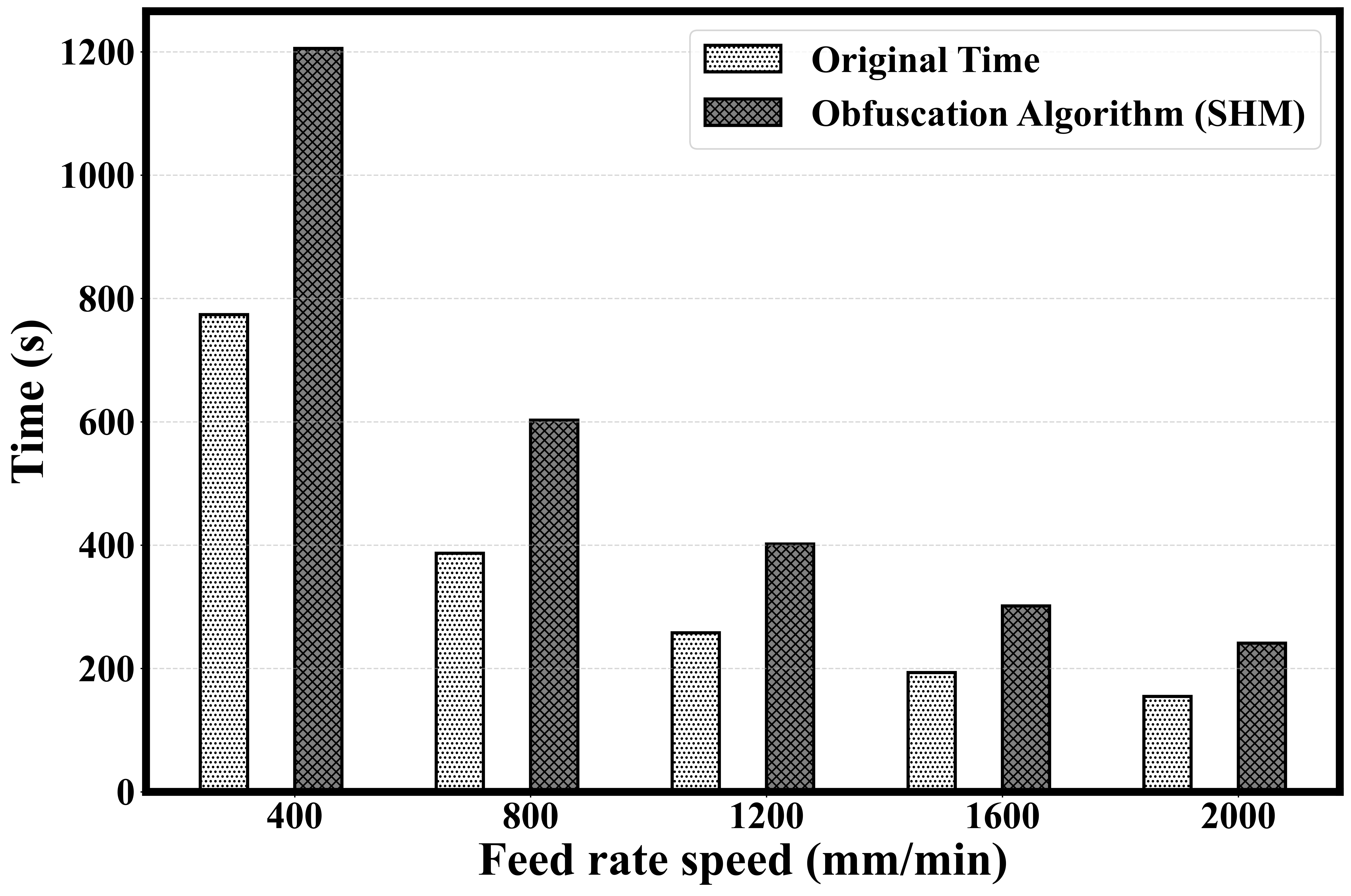}
    \caption{Comparing the timing overhead of the SHM algorithm at different printing speeds for the key speciemen.}
    \label{fig:printtime}
\end{figure}

It is worth noting that both of our metrics \( A \) and \( B \) are dimensionless as A is the sum of points and D is the minimum sum of the squared error. As shown in Algorithm \ref{alg:apprhac2alg}, we first obtain the original shape and its convex hull. Then, we add random rectangles to random locations under two conditions. First, the rectangle must intersect with the convex hull and overlap with the original shape. Second, the newly added rectangle must be connected to the original shape to avoid creating isolated components. After each addition, we record the area and dissimilarity. This process continues until the area of the modified shape exceeds that of the convex hull. From the recorded data, we select the shape that has the highest reward as the optimized shape. We also apply a binary closing algorithm to fill gaps and holes within the shape. Although this algorithm works well for concave shapes, it may not be practical for convex shapes. For instance, the convex hull of a rectangle or a triangle is the shape itself. In such cases, we first enclose the shape within a larger bounding rectangle by randomly extending its length and width. Then, instead of adding points, we remove random points from the area between the bounding rectangle and the original shape. The result of our algorithm is shown in Figure \ref{fig:before_after}. As it is shown in the reward plot, at some point there is a maximum peak in the reward function where the added area is low while the dissimilarity is high enough. As an example, we applied our combined method to obfuscate the key object shown in Figure~\ref{fig:before_after}. As shown in the figure the important information about the teeth of the key is successfully obfuscated using the SHM Approach. The normal print time for this three-layer object was approximately 258 seconds. SHM Approach added around 143 seconds, resulting in a 55\% increase in print time. We also provide a  comparison of the printing time for the key example, as shown in Figure~\ref{fig:printtime}. The printing times were calculated based on different feed rates. Feed rate refers to the speed at which the nozzle moves. As shown in  Figure~\ref{fig:printtime}, at lower speeds. SHM Approach consistently introduces a linear overhead regardless of speed, as the total path length remains constant and no recovery time is needed.

\begin{algorithm}[!t]
\caption{SHM algorithm}\label{alg:apprhac2alg}
\begin{algorithmic}[1]
\Function{AddRects}{$img,\ img\_neg,\ n, $}
    \State $out \gets img$
    \State $(h, w) \gets \Call{Shape}{img}$
    \For{$i = 1$ to $n$}
        \For{$j = 1$ to $attempt$}
            \State $rw \gets \Call{RandInt}{min\_s, max\_s}$
            \State $rh \gets \Call{RandInt}{min\_s, max\_s}$
            \State $x \gets \Call{RandInt}{0, w - rw}$
            \State $y \gets \Call{RandInt}{0, h - rh}$
            \State $r \gets out[y{:}y{+}rh,\ x{:}x{+}rw]$
            \State $r2 \gets img\_neg[y{:}y{+}rh,\ x{:}x{+}rw]$
            \If{$(\Call{Sum}{r2} \land \Call{Sum}{r})\ $}
                \State $out[y{:}y{+}rh,\ x{:}x{+}rw] \gets 1$
                \State \textbf{break}
            \EndIf


        \EndFor
    \EndFor
    \State \Return $out$
\EndFunction

\State $convex\_mask \gets \Call{convex\_hull}{img}$
\State $img\_neg \gets \ \sim (convex\_mask\ \&\ img\_points)$

\State $prev\_i \gets -1$
\For{$i = Start$ to $Stop$ step $Step$}
    \If{$prev\_i < 0$}
        \State $prev\_i \gets Start - 100$
    \EndIf
    \State $n \gets i - prev\_i$
    \State $result\_img \gets \Call{AddRects}{result\_img,\ img\_neg,\ n}$
    \State $result\_img \gets \Call{BinaryClosing}{result\_img}$
    \State $area\_img \gets \Call{Sum}{result\_img}$
    \State $area\_convex \gets \Call{Sum}{convex\_mask}$
    
    \If{$area\_img > area\_convex$}
        \State \textbf{break}
    \EndIf
    \State $disp \gets \Call{Procrustes}{result\_img,\ img\_points}$
    \State $r2 \gets result\_img\ \&\ img\_neg$
    \State $area \gets \Call{Sum}{r2}$
    \State \Call{Append}{$all\_score,\ disp$}
    \State \Call{Append}{$all\_area,\ area$}
    \State \Call{Append}{$all\_shapes,\ result\_img$}
    \State $prev\_i \gets i$
\EndFor
\State $reward \gets all\_area-all\_score$
\State $Optimized\_idx \gets arg\max(reward)$
\State $Optimized\_shape \gets all\_shapes[Optimized\_idx]$

\end{algorithmic}
\end{algorithm}


  \subsection{Evaluating the proposed obfuscation method}
To evaluate our method, we reused the earlier triangle example, but this time enclosed the shape in a rectangular polygon (Figure \ref{fig:approach2}(a)) to mask fan noise and sudden acoustic spikes. After recording and preprocessing the audio, we detected the transition points and reconstructed the object. The attacker would recover only the outer rectangle (Figure \ref{fig:approach2}(b)), not the original triangle. This shows that, even with full audio access, the original object remains hidden.
The optimized SHM result is shown in Figure \ref{fig:before_after}. We used this optimized polygon for obfuscation, printed the object, recorded the audio, and the attacker’s reconstructed output appears in Figure \ref{fig:before_after}.

\section{Conclusion}
\label{sec:conclusion}
In this work, we first investigated the impact of cooling fan and stepper motor noise on data leakage. We then employed a spike detection method to reconstruct the baseline shape. Next, we proposed the Stealth Head Movement (SHM) obfuscation algorithm to conceal motion patterns. This approach ensures that even if an attacker detects sudden movements or direction changes in the stepper motors, he still cannot reconstruct the full object. Subsequently, we formulated our method and provided an optimized solution to minimize print time. The main advantage of our approach is that it requires no additional hardware and is fully compatible with existing 3D printers. Implementation involves only minimal modifications to the G-code file to incorporate the SHM algorithm, eliminating the need for any extra setup such as noise-canceling or noise-generating devices.


\begin{acks}
This work was supported by the NSF Grant CCRI 2234972. Any
opinions, conclusions, findings, or recommendations provided in
this work do not necessarily reflect the views of the funding
organizations.
\end{acks}


\bibliographystyle{ACM-Reference-Format}

\end{document}